\documentclass[twocolumn]{jpsj3}
\usepackage[final]{changes}
\usepackage{txfonts}
\usepackage{bm}

\title{Magnetic Properties under Pressure in Novel Spin-Triplet Superconductor UTe$_2$}

\author{
Dexin~Li$^1$,
Ai~Nakamura$^1$,
Fuminori~Honda$^{1,2}$,
Yoshiki~J.~Sato$^1$,
Yoshiya~Homma$^1$,
Yusei~Shimizu$^1$,
Jun~Ishizuka$^3$,
Youichi~Yanase$^{4,5}$,
Georg~Knebel$^6$,
Jacques~Flouquet$^6$, and
Dai~Aoki$^{1,6}$\thanks{aoki@imr.tohoku.ac.jp}
}

\inst{%
$^1$IMR, Tohoku University, Oarai, Ibaraki, 311-1313, Japan\\
$^2$Central Institute of Radioisotope Science and Safety Management, Kyushu University, Fukuoka, 819-0395, Japan\\
$^3$Institute for Theoretical Physics, ETH Zurich, 8093 Zurich, Switzerland\\
$^4$Department of Physics, Graduate School of Science, Kyoto University, Kyoto 606-8502, Japan\\
$^5$Institute for Molecular Science, Okazaki 444-8585, Japan\\
$^6$Univ. Grenoble Alpes, CEA, Grenoble INP, IRIG, PHELIQS, F-38000 Grenoble, France\\
}

\abst{%
We report the magnetic susceptibility and the magnetization under pressures up to $1.7\,{\rm GPa}$ above the critical pressure, $P_{\rm c}\sim 1.5\,{\rm GPa}$, for $H\parallel a$, $b$ and $c$-axes in the novel spin triplet superconductor UTe$_2$.
The anisotropic magnetic susceptibility at low pressure with the easy magnetization $a$-axis changes to the quasi-isotropic behavior at high pressure, 
revealing a rapid suppression of the susceptibility for $a$-axis, and a gradual increase of the susceptibility for the $b$-axis.
At $1.7\,{\rm GPa}$ above $P_{\rm c}$, magnetic anomalies are detected at $T_{\rm MO}\sim 3\,{\rm K}$ and $T_{\rm WMO}\sim 10\,{\rm K}$.
The former anomaly corresponds to long-range magnetic order, most likely antiferromagnetism,
while the latter shows a broad anomaly, which is probably due to the development of short range order.
The unusual decrease and increase of the susceptibility below $T_{\rm WMO}$ for $H\parallel a$ and $b$-axes, respectively, 
indicate the complex magnetic properties at low temperatures above $P_{\rm c}$.
This is related to the interplay between multiple fluctuations dominated by antiferromagnetism, ferroamgnetism, valence and Fermi surface instabilities.
}

\begin{document}
\maketitle
%====================================
The observation of unconventional superconductivity in UTe$_2$ is one of the most interesting topics in heavy-fermion physics.~\cite{Ran19,Aok19}
UTe$_2$ crystallizes a body centered orthorhombic structure with the space group $Immm$ (No. 71, $D_{2h}^{25}$). 
No long range magnetic order is detected down to low temperatures and the ground state is paramagnetic.~\cite{Sun19,Pau20}
The Sommerfeld coefficient of the specific heat is $\gamma \approx 120\,{\rm mJ K^{-2}mol^{-1}}$, revealing a heavy electronic state,
which is also demonstrated by a typical heavy-fermion behavior in electrical resistivity with the large $A$ coefficient for the $T^2$ dependence below $5\,{\rm K}$.
Superconductivity is observed below $T_{\rm c}=1.6\,{\rm K}$ associated with a large specific heat jump indicating the strong coupling nature.

One of the most remarkable features in UTe$_2$ is the huge superconducting upper critical field $H_{\rm c2}$.
For $H\parallel b$-axis, field-reentrant superconductivity is observed up to $H_{\rm m}\approx 35\,{\rm T}$, 
at which a first order metamagnetic transition occurs.~\cite{Kne19,Ran19_HighField}
Superconductivity is cutoff above $H_{\rm m}$ probably due to a drastic change of the electronic state.\cite{Niu20}
It is pointed out that the huge $H_{\rm c2}$ displaying the reentrant behavior resembles those observed in ferromagnetic superconductors, URhGe~\cite{Aok01} and UCoGe~\cite{Huy07},
where field-reentrant (-reinforced) superconductivity is induced due to the development of ferromagnetic fluctuations associated with the collapse of the Curie temperature $T_{\rm Curie}$.~\cite{Lev07,Aok09_UCoGe,Aok12_JPSJ_review,Aok19}
However, UTe$_2$ is a paramagnet and the values of $H_{\rm c2}$ are large in all field directions. 
Ferromagnetic fluctuations in $\mu$SR~\cite{Sun19} and longitudinal magnetic fluctuations in NMR~\cite{Tok19} are reported.
On the other hand, inelastic neutron experiments reveal incommensurate antiferromagnetic excitations~\cite{Dua20}.
Thus, the magnetic properties and fluctuations in UTe$_2$ are more complicated than those in ferromagnetic superconductors,
although spin-triplet superconductivity is most likely realized in both systems. 
Strongest evidence for spin-triplet superconductivity in UTe$_2$ was given by the Knight shift of NMR experiments,
which shows only a tiny decrease of the spin susceptibility below $T_{\rm c}$.~\cite{Nak19,Nak20}

Another important point in UTe$_2$ is the appearance of multiple superconducting phases under pressure.~\cite{Bra19}
The superconducting transition at $T_{\rm c}$ splits into two above $0.25\,{\rm GPa}$.
The lower $T_{\rm c}$ decreases continuously with pressure, while the higher $T_{\rm c}$ takes a maximum of $3\,{\rm K}$ at $\sim 1\,{\rm GPa}$, and decreases rapidly at higher pressures.
The split of $T_{\rm c}$ under pressure gives rise to the unusual $H_{\rm c2}$ curve under pressure. 
For $H \parallel a$-axis, $H_{\rm c2} (T)$ shows an abrupt enhancement at low temperature, which is connected to the lower $T_{\rm c}$ at zero field,
revealing the occurence of multiple superconducting phases under magnetic field as well.~\cite{Aok20_UTe2}
Above the critical pressure, $P_{\rm c}\sim 1.5\,{\rm GPa}$, superconductivity is suppressed and a magnetically ordered state appears.
Interestingly, just above $P_{\rm c}$, field-induced superconductivity is detected far above $20\,{\rm T}$ in the spin-polarized state above the critical field, where the magnetic ordered state is suppressed.~\cite{Aok21_UTe2}

In order to clarify the magnetic properties under pressure, we performed magnetization and magnetic susceptibility measurements up to pressures above $P_{\rm c}$ for $H\parallel a$, $b$ and $c$-axes. 
We show that the magnetic anisotropy changes with pressure and the magnetic easy-axis is reversed at low temperature between $a$ and $b$-axes. 
Two magnetic anomalies were clearly detected above $P_{\rm c}$.

Single crystals of UTe$_2$ were grown using the chemical vapor transport method as described elsewhere~\cite{Aok20_SCES}.
The magnetic susceptibility and the magnetization measurements under pressure were performed at temperatures down to $2\,{\rm K}$ and at fields up to $5\,{\rm T}$ using a piston cylinder cell up to $1.7\,{\rm GPa}$ in a commercial SQUID magnetometer.~\cite{sup}

%========================================================================================
\begin{fullfigure}[bth]
\begin{center}
\includegraphics[width=\hsize]{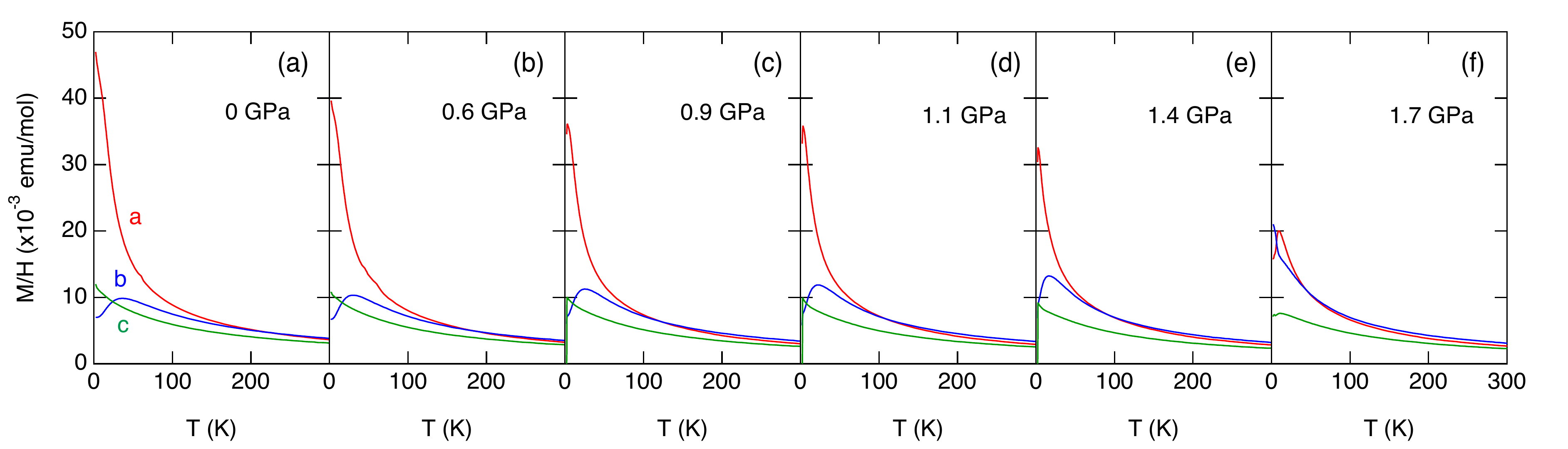}
\end{center}
\caption{(Color online) Magnetic susceptibilities, $\chi_a$, $\chi_b$ and $\chi_c$ for $H\parallel a$, $b$ and $c$-axes, respectively at $1\,{\rm T}$ at different pressures, 0, 0.6, 0.9, 1.1, 1.4 and $1.7\,{\rm GPa}$ in UTe$_2$.}
\label{fig:sus_all}
\end{fullfigure}
%========================================================================================
Figure~\ref{fig:sus_all} shows the magnetic susceptibilities $M/H$ for $H \parallel a$, $b$, and $c$-axes ($\chi_a$, $\chi_b$, and $\chi_c$) at $1\,{\rm T}$ at different pressures.
It is obvious that the magnetic anisotropy can be tuned by the external parameters. 
Already at ambient pressure the easy-magnetization axis switches from the $a$ to the $b$-axis under magnetic field at the metamagnetic transition at $H_{\rm m}=35\,{\rm T}$.
At ambient pressure, the $a$-axis is the easy-magnetization axis at low field, revealing the rapid increase of $\chi_a$ on cooling without the saturation down to $T_{\rm c}$.
$\chi_b$ shows a broad maximum around $T_{\chi,\rm max}\sim 35\,{\rm K}$, which is connected to the first order metamagnetic transition $H_{\rm m}=35\,{\rm T}$ at low temperature. 
It should be noted that the anisotropy between $\chi_a$ and $\chi_b$ is small at high temperatures above $150\,{\rm K}$ contrary to the large anisotropy at low temperatures,
where $\chi_a$ is by a factor 7 higher than $\chi_b$ at ambient pressure. 

Increasing pressure leads to an decrease of $\chi_a$,
while $\chi_b$ increases with pressure, as shown in Fig.~\ref{fig:sus_all}. 
These results are consistent with the prediction through the Maxwell relation by magnetostriction measurements~\cite{Tho21}.
$\chi_c$ is slightly reduced with pressure.
These different pressure responses give rise to the change of magnetic anisotropy by applying the pressure.
The susceptibility ratio between $a$, $b$ and $c$-axes represented by $\chi_a/\chi_b$, $\chi_a/\chi_c$ and $\chi_b/\chi_c$~\cite{sup1} also indicates that
the magnetic anisotropy becomes small with pressure, retaining the hard-magnetization $c$-axis.
At $1.7\,{\rm GPa}$ ($> P_{\rm c}$), the $b$-axis becomes the easy-magnetization axis at low temperature.
The decrease of $\chi_a$ seems to be driven by the creation of an antiferromagnetic molecular field.
Above $P_{\rm c}$ it is speculated that the antiferromagnetic moment might be aligned along the $a$-axis by an exchange antiferromagnetic molecular field from the fact that the magnetization for $H\parallel a$-axis is slightly smaller than that for $H\parallel b$-axis.
Note that recent magnetostriction experiments along the $c$-axis for $H\parallel a$, $b$ and $c$-axes predict
a switching of the easy-magnetization axis between $a$ and $b$-axis under uniaxial stress along $c$-axis.~\cite{Wil21}

%========================================================================================
\begin{figure}[tbh]
\begin{center}
\includegraphics[width= 0.8\hsize,clip]{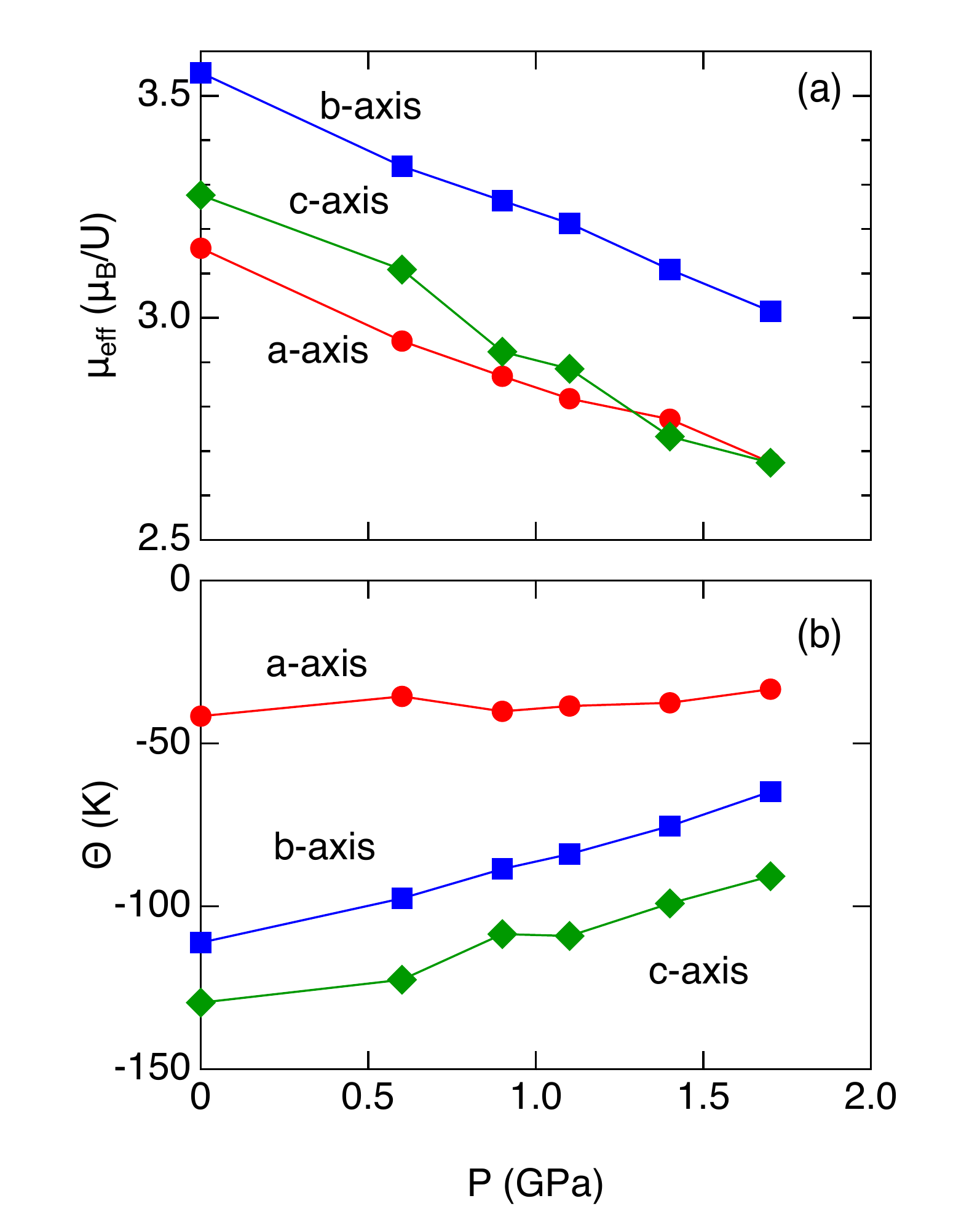}
\end{center}
\caption{(Color online) Pressure dependence of (a) the effective moments and (b) the Weiss constants for $H\parallel a$, $b$, and $c$ in UTe$_2$.}
\label{fig:mu_eff_weiss}
\end{figure}
%========================================================================================
The change from anisotropic to quasi-isotropic magnetic properties is also seen in the pressure dependence of the Weiss constants, $\Theta$.
The inverse magnetic susceptibilities follow a Curie-Weiss behavior $\chi = C/(T-\Theta)$  at high temperatures for $a$, $b$ and $c$-axes in all pressure range.~\cite{sup2} 
A linear fitting to the inverse magnetic susceptibility from $300$ to $150\,{\rm K}$ yields the effective moment $\mu_{\rm eff}$ and the Weiss constant $\Theta$, as shown in Fig.~\ref{fig:mu_eff_weiss}.
The Weiss constants are always negative, indicating that antiferromagnetic correlations are dominant in first approximation. 
Increasing pressure, the Weiss constants for $a$-axis are almost the same, 
whereas the Weiss constants for $b$ and $c$-axes increase, getting close to the value of the $a$-axis;
this indicates that the magnetic anisotropy becomes small with increasing the pressure at least at high temperature region.

The pressure evolution of the effective moments are also derived from the fitting, as shown in Fig.~\ref{fig:mu_eff_weiss}(a).
At ambient pressure, the effective moments are ranging from $3.55$ to $3.14\,\mu_{\rm B}$,
which are close to the free ion values, $3.58\,\mu_{\rm B}$ and $3.62\,\mu_{\rm B}$ for 5$f^2$ (U$^{4+}$) or 5$f^3$ (U$^{3+}$).
The effective moments decrease with pressure, suggesting the deviation from the 5$f^2$ or 5$f^3$ configurations,
or combined effects of crystalline electric field and antiferromagnetic correlations.

%========================================================================================
\begin{figure}[tbh]
\begin{center}
\includegraphics[width= 0.7\hsize,clip]{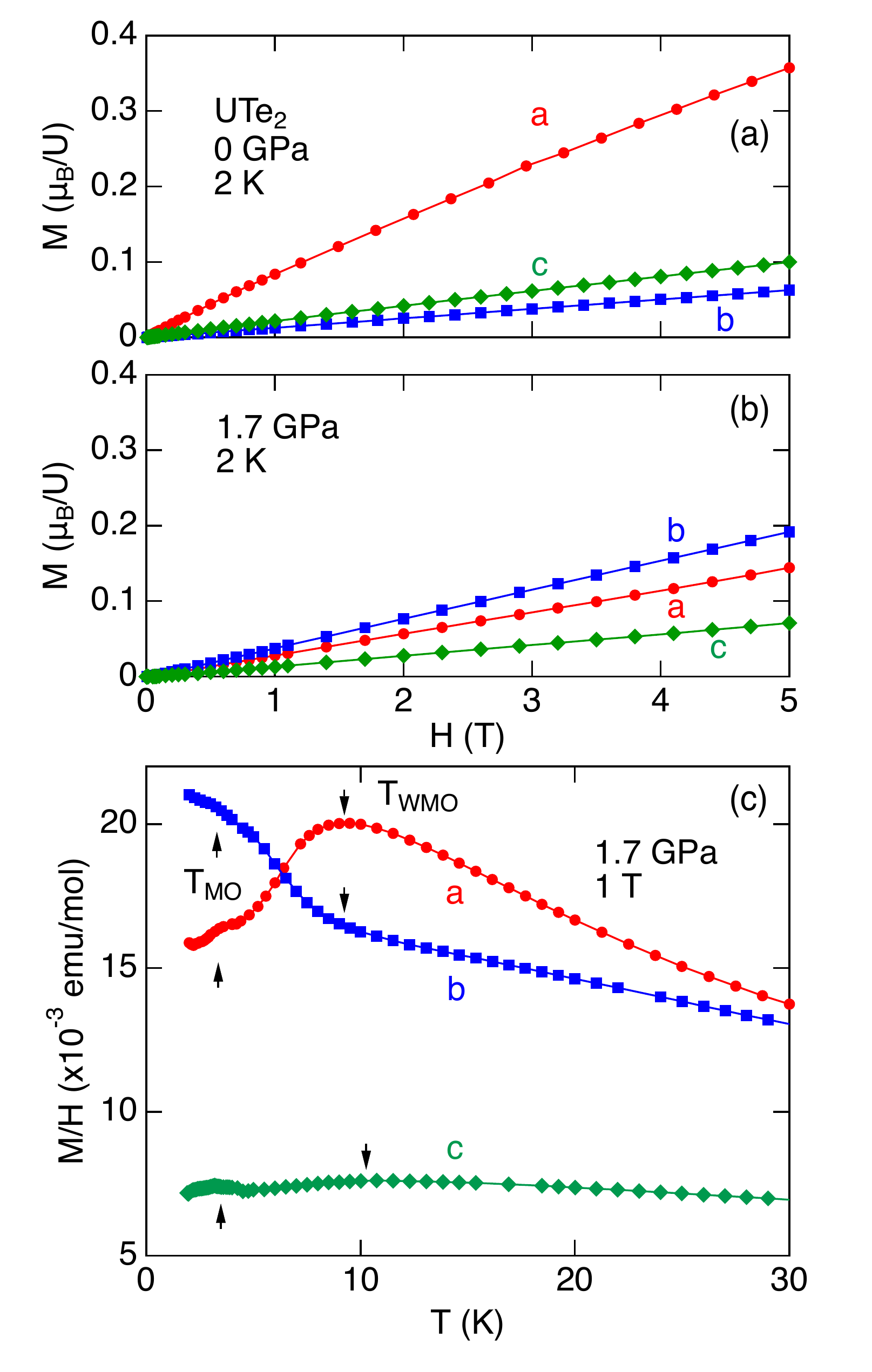}
\end{center}
\caption{(Color online) Magnetization curves at $2\,{\rm K}$ for $H\parallel a$, $b$ and $c$-axes at (a) ambient pressure and (b) $1.7\,{\rm GPa}$. (c) Temperature dependence of the magnetic susceptibilities at $1.7\,{\rm GPa}$ for $H\parallel a$, $b$ and $c$-axes.}
\label{fig:sus_highP}
\end{figure}
%========================================================================================
At $1.7\,{\rm GPa}$, above the critical pressure $P_{\rm c}$,
magnetic anomalies are clearly detected at $T_{\rm WMO}$ and $T_{\rm MO}$ in $\chi_a$, $\chi_b$ and $\chi_c$, as shown in Fig.~\ref{fig:sus_highP}(c).
$\chi_a$ shows a broad maximum at $T_{\rm WMO}\approx 9.5\,{\rm K}$.
$\chi_c$ also displays a small and broad maximum at $T_{\rm WMO} \approx 10.5\,{\rm K}$.
In difference, $\chi_b$ shows the gradual increase through $T_{\rm WMO}$ on cooling, and clearly, the easy magnetization axis changes from $a$ to $b$-axis.
Since the anomaly at $T_{\rm WMO}$ is very broad with a crossover-like behavior, it cannot be defined as a clear appearance of the long range magnetic order.
The increase of $\chi_b$ below $T_{\rm WMO}$ may reflect the strong feedback between antiferromagnetic and ferromagnetic interactions with a far weaker effect than that at ambient pressure, but seems to be due to the particular crystal structure of UTe$_2$.
If the anomaly at $T_{\rm WMO}$ corresponds to a ferromagnetic order, the anomaly should be immediately smeared out at high fields for $b$-axis.
However, $T_{\rm WMO}$ is still visible up to $5\,{\rm T}$, indicating that it does not correspond to the ferromagnetism.
Note that $T_{\rm WMO}$ is also observed in resistivity and AC calorimetry measurements, but the anomaly due to $T_{\rm WMO}$ is rather broad at low fields.

At low temperature, another magnetic anomaly is detected around $3\,{\rm K}$, as denoted by $T_{\rm MO}$ in Fig.~\ref{fig:sus_highP}(c).
The anomaly at $T_{\rm MO}$ is much sharper than that of $T_{\rm WMO}$,
as it is also detected more clearly in resistivity and AC calorimetry.\cite{Aok21_UTe2}
It seems obvious that $T_{\rm MO}$ corresponds to long range magnetic order, most likely antiferromagnetism.\cite{Bra19, Aok20_UTe2, Tho20}
At least one can say that no evidence for ferromagnetic order appears at $T_{\rm MO}$ and $T_{\rm WMO}$.

Figures~\ref{fig:sus_highP}(a)(b) show the comparison of magnetizations at ambient pressure and at $1.7\,{\rm GPa}$, above the critical pressure $P_{\rm c}$ at $2\,{\rm K}$. 
The $b$-axis is clearly above $P_{\rm c}$ the easy magnetization axis, and no more the $a$-axis.
The anisotropy is rather small for the $a$, $b$ and $c$-axes . 
This is clear contrast to ambient pressure, where the easy $a$-axis reveals more Ising-like behavior.

%========================================================================================
\begin{figure}[tbh]
\begin{center}
\includegraphics[width= 0.7\hsize,clip]{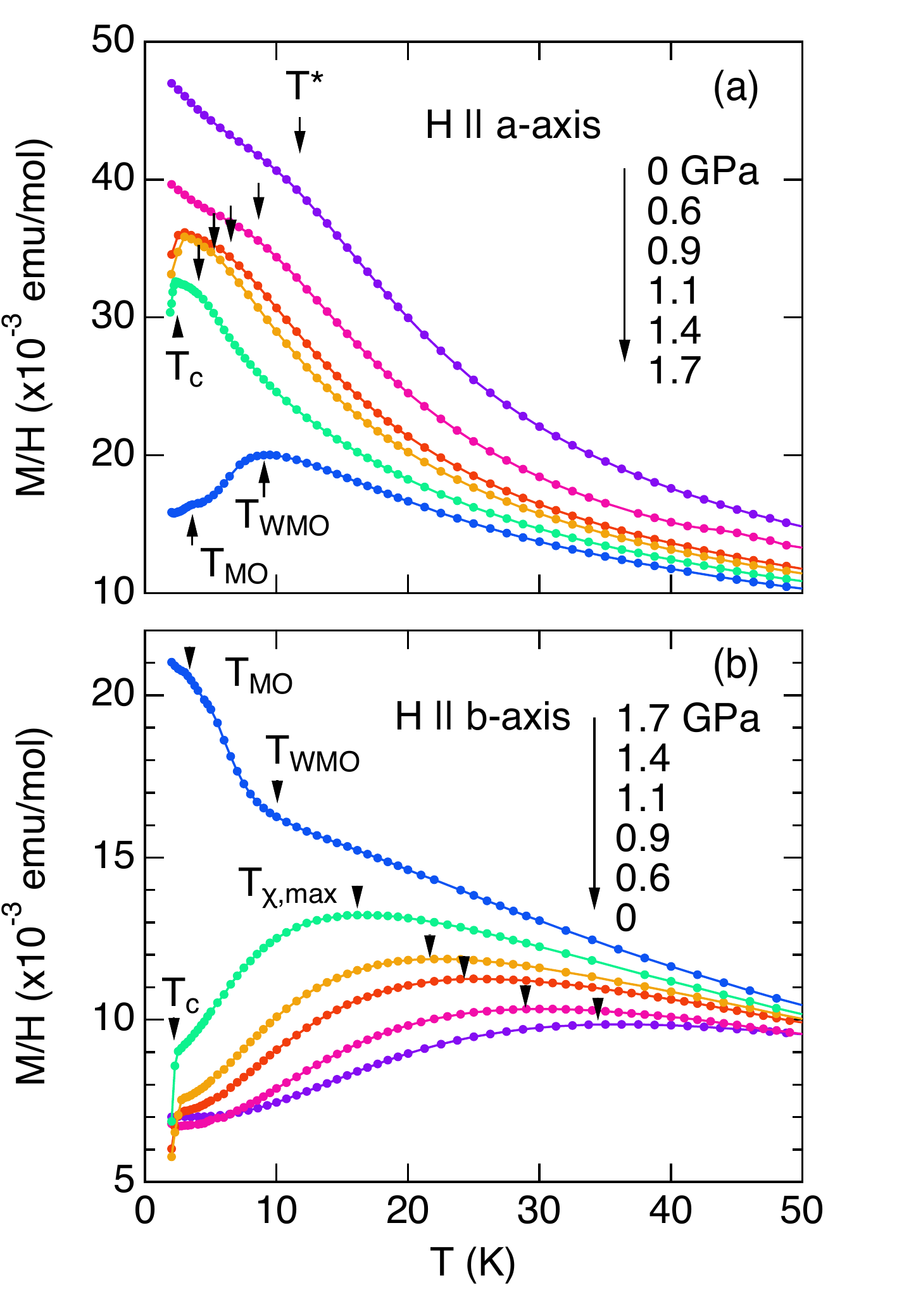}
\end{center}
\caption{(Color online) Temperature dependence of the magnetic susceptibilities measured at $1\,{\rm T}$ for $H\parallel a$ and $b$-axes at different pressure.}
\label{fig:sus_ab}
\end{figure}
%========================================================================================
Figure~\ref{fig:sus_ab} shows the temperature dependence of the susceptibility at low temperatures for $H\parallel a$ and $b$-axes at different pressures.
For $H\parallel a$-axis, a broad shoulder appears around $T^\ast\sim 12\,{\rm K}$, which shifts to the lower temperatures with increasing pressure. 
Correspondingly, the sharp increase of the susceptibility on cooling at ambient pressure tends to be suppressed with pressure. 
At pressures in the range from $0.9$ to $1.4\,{\rm GPa}$, 
the susceptibility drops rapidly below $\sim 3\,{\rm K}$ due to the onset of superconductivity.

For $H\parallel b$-axis, a broad maximum of the susceptibility around $T_{\chi,\rm max}\sim 35\,{\rm K}$ becomes sharp and shifts to the lower temperature with increasing pressure. The decrease of $T_{\chi,\rm max}$ is well scaled with the decrease of the metamagnetic field $H_{\rm m}$ under pressure.~\cite{Kne20}
Above $P_{\rm c}$ the maximum has vanished and the magnetic state develops below $T_{\rm WMO}$. 

\begin{figure}[tbh]
\begin{center}
\includegraphics[width= 0.8\hsize,clip]{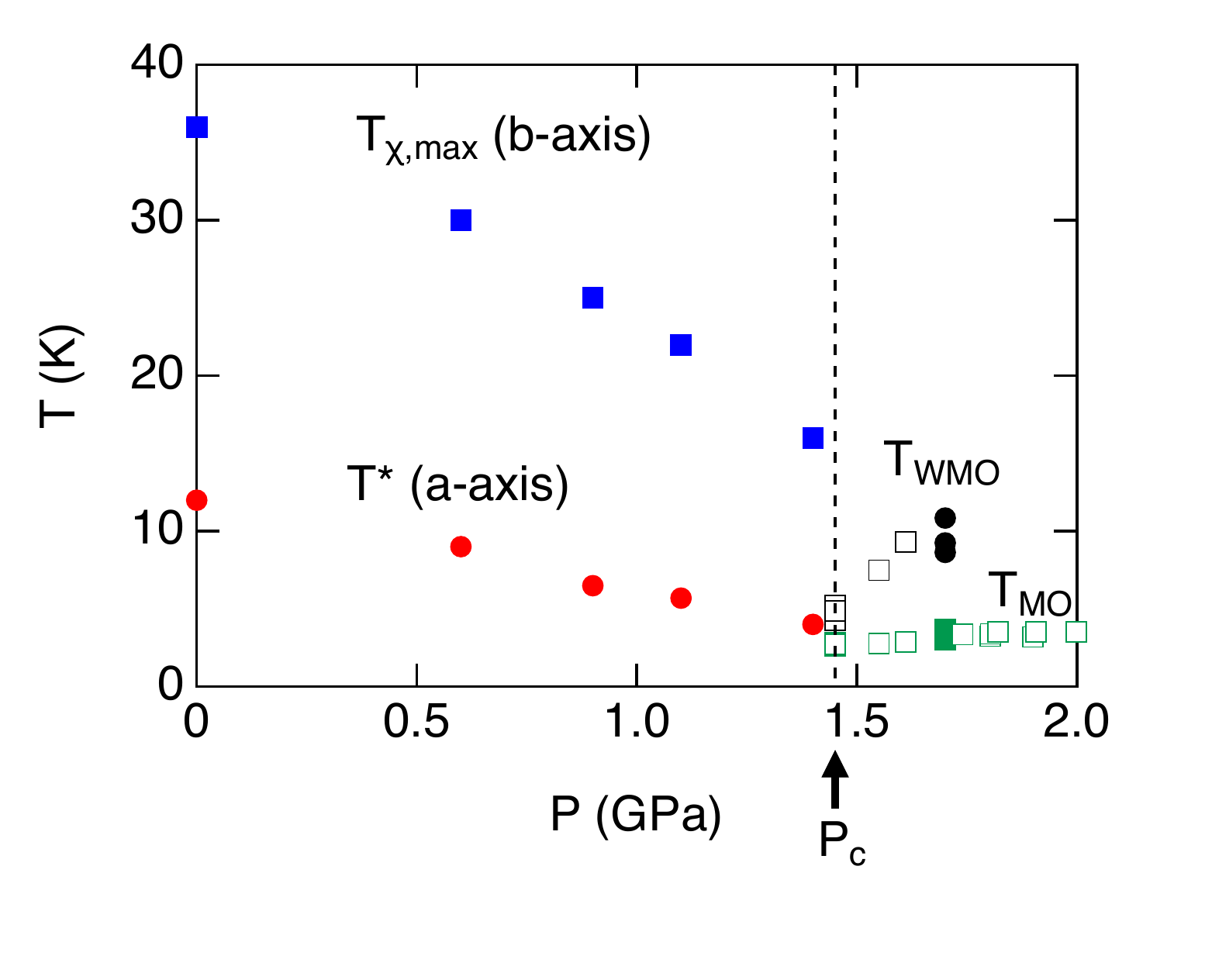}
\end{center}
\caption{(Color online) Pressure dependence of $T_{\chi,\rm max}$ and $T^\ast$ at $1\,{\rm T}$ for $H\parallel a$ and $b$-axes, respectively. $T_{\rm WMO}$ and $T_{\rm MO}$ at $1\,{\rm T}$ are also plotted. The open symbols for $T_{\rm WMO}$ and $T_{\rm MO}$ are the results of resistivity and AC calorimetry measurements at zero field cited from Ref.~\citen{Bra19,Aok20_UTe2,Aok21_UTe2}.}
\label{fig:T0_Tchimax}
\end{figure}

Figure~\ref{fig:T0_Tchimax} shows the pressure dependence of $T^\ast$ and $T_{\chi,\rm max}$ together with $T_{\rm WMO}$ and $T_{\rm MO}$ above $P_{\rm c}$.
Both $T_{\chi,\rm max}$ and $T^\ast$ decrease with increasing pressure, and are supposed to become zero around $2\,{\rm GPa}$.
But they have finite values just below $P_{\rm c}$ and abruptly disappear;
this implies that the magnetic properties and electronic states drastically change at $P_{\rm c}$ with the first order transition.
$T_{\rm MO}$ and $T_{\rm WMO}$ detected by susceptibility measurements are consistent with those detected by resistivity and AC calorimetry measurements.
Note that $T_{\rm MO}$ and $T_{\rm WMO}$ are not connected continuously, revealing the appearance of the first order transition at $P_{\rm c}$ again.
The slope for $T^\ast$ is $\partial T^\ast/\partial P \sim -5.8\,{\rm K/GPa}$, which is in excellent agreement with the value of $-6\,{\rm K/GPa}$ estimated from the thermal expansion and the electronic Gr\"{u}neisen parameter, $\Gamma_{\rm e}=-30$.~\cite{Wil21,Tho21}
The decrease of $T^\ast$ is also consistent with the results of resistivity measurements under pressure~\cite{Ran20_pressure}.

%========================================================================================
%========================================================================================

The rather high compressibility of UTe$_2$ leads to a volume reduction close to 3 {\%} at 1.7 GPa,~\cite{Hon21} 
which is a large volume change in condensed matter. 
For example, nearly the same volume reduction in $^3$He ($\sim 5\,{\%}$) occurs, 
where the paramagnetic Fermi liquid phase~\cite{Leg75} changes to the solid phase on the melting curve at very low temperature.~\cite{Mil61}
It is far from ferromagnetic instability~\cite{Flo82}, and phase transitions to the spin triplet superfluid phases A and B occur on cooling.
Further on cooling, the antiferromagnetic order also develops in the solid phase in the millikelvin range.~\cite{Osh80,Ben85}
Many examples for the large volume reduction have been found in the heavy fermion systems~\cite{Flo82} as well as in a ladder system with a few {\%} of the volume reduction~\cite{Iso98}, where drastic changes of electronic and magnetic properties were observed.
Indeed, in UTe$_2$, a substantial shrink in the volume may cause the Lifshitz transition under the pressure, as indicated by a first-principles band calculation based on the GGA+$U$ method. 
As shown in the supplemental material~\cite{sup3}, a Fermi surface may appear around the $\Gamma$ point in the Brillouin zone, which has not been found in calculations at ambient pressure.~\cite{Ish19,Xu19} 
It may affect the magnetic properties.

The particularity in U compounds is the switch from a normalization of the wave function to the U$^{3+}$ configuration toward the U$^{4+}$ configuration (see similar discussion on Tm and Sm intermediate valence compounds for Tm$^{2+}$/Tm$^{3+}$ and Sm$^{2+}$/Sm$^{3+}$, respectively in ref.~\citen{Der06}). 
On the other hand, the situation is very different from that in Ce compounds~\cite{Flo09}, where valence fluctuations occur between a single occupied 4$f^1$ electron for Ce$^{3+}$ and and an empty 4$f$ shell for Ce$^{4+}$;
while in Yb compounds the situation correspond to the case of one hole in 4$f$ shell (Yb$^{3+}$) and a full occupancy (Yb$^{2+}$). 
In Ce compounds, starting with a nearly trivalent valence, the pressure will generally increase the Kondo interaction and thus the Kondo temperature $T_{\rm K}$ increases; 
when $T_{\rm K}$ reaches the crystal field value $\Delta_{\rm cf}/k_{\rm B}$ ($\Delta_{\rm cf}$: the crystal field splitting energy) this last one is no more efficient to lift the six fold degeneracy of the total angular momentum $J=5/2$.~\cite{Flo09} 
Recently, the dual nature of 5$f$ electrons in U compounds is nicely proved microscopically by high energy spectroscopy experiments~\cite{Amo20}. 
A small valence change under pressure in UTe$_2$ has already been emphasized~\cite{Aok20_UTe2} and directly shown by x-ray spectroscopy~\cite{Tho20}.
A strong indication for the mixed valence state with a dominant $5f^3$ (U$^{3+}$) configuration has been recently reported in a Core-level spectroscopy study~\cite{Fujimori21}. 
The interplay of the  U$^{3+}$ and U$^{4+}$ configurations has been also observed in high resolution angle resolved photoemission spectroscopy.~\cite{Miao20}

A supplementary ingredient is that UTe$_2$ forms a unique orthorhombic structure with the space group $Immm$, which cannot be seen in other dichalcogenide family, such as USe$_2$ and US$_2$.
In UTe$_2$ the two U chains form a two-leg ladder along the $a$-axis.
Thus, the possibility of strong magnetic frustrations and the absence of long-range orders are pointed out~\cite{Xu19}.
The unique crystal structure of UTe$_2$ leads to a singular band structure emerging from a Kondo semiconductor~\cite{Har20} at the limit of vanishing correlation linked to the weakness of the Coulomb repulsion $U$ in regards to the transfer hopping integral $p$ in reference~\cite{Ish21}. 
As observed in SmB$_6$~\cite{Bar05} or SmS~\cite{Bar04}, pressure will add another channel to close the Kondo gap .

A remarkable point is that crossing $P_{\rm c}$ the magnetic anisotropy between $a$ and $b$-axes collapses and is even reversed in its sign. 
At ambient pressure, a ratio $\chi_a/\chi_b$ of 7 just above $T_{\rm c}$ has been taken as a possible classification of UTe$_2$ has an Ising system. 
The pressure has led to a full undressing of the initial crystal field splitting i. e. in the selection between the $j=5/2$ and $j=7/2$ angular component of the 5$f$ electrons. 
It was already stressed in a theoretical model~\cite{Ish21} that the strength of $U/p$ ($U$: Coulomb repulsion, $p$: hopping integral) will push towards an Ising limit for the spin and that pressure by lowering $U/p$ will weaken the Ising character. 
In this model no direct link was made between a drastic or slightly change of valence through $P_{\rm c}$ and a deep modification of the crystal field. 
In ref.\citen{Ish21} it has been argued that the pressure will lower the strength of the Coulomb repulsion $U$ and thus modify the strength of the ferromagnetic and antiferromagnetic interactions, as observed in experiment. 
Despite the Ising character of ferromagnetic fluctuation, the antiferromagnetic fluctuations are predicted to show a moderately quasi-isotropic property (Heisenberg character).
In general, the Ising ferromagnetic fluctuations are favorable for spin-triplet superconductivity,
while the antiferromagnetic fluctuations are favorable for the spin-singlet state.
The change to the quasi-isotropic properties with pressure suggests the development of the antiferromagnetic fluctuations, implying the parity mixing.
This is also inferred from the experimental results of multiple superconducting phases and $H_{\rm c2}(T)$ curves under pressure.~\cite{Bra19,Aok20_UTe2,Kne20}

In the trivalent U$^{3+}$ with the 5$f^3$ configuration, the crystal field level scheme leads to the series of five Kramers doublets for $J=9/2$ in the orthorhombic structure, which gives rise to heavy fermions properties quite similar to those of Ce$^{3+}$ or Yb$^{3+}$. 
On the other hand, the tetravalent U$^{4+}$ with the 5$f^2$ configuration will lead to the nine singlets for $J=4$, which may leave the possibility of deep change of the crystal field and thus of the magnetic anisotropy. 
For the 4$f$ systems it is equivalent of the 4$f^2$ configuration, such  as Pr$^{3+}$. 
For UTe$_2$, the dual localized and itinerant character of the 5$f$ electrons~\cite{Amo20,Zwi03} seems the origin of the observed drastic change of the magnetic anisotropy with pressure; 
at high temperature signature of the localized character is marked by the nice Curie-Weiss behaviors in the all pressure range.

In summary, our experiments reveal that the anisotropic magnetic susceptibility at ambient pressure in UTe$_2$ changes to the quasi-isotropic behavior under pressure.
Above $P_{\rm c}$, two magnetic anomalies are detected at $T_{\rm MO}$ corresponding to the antiferromagnetic order and $T_{\rm WMO}$, which is most likely a crossover rather than a static magnetic order, revealing the complex magnetic properties.
These pressure sensitive magnetic properties are probably due to the unique crystal structure associated with the strong volume reduction, ferromagnetic/antiferromagnetic fluctuations, valence and Fermi surface instabilities.

\section*{Acknowledgements}
We thank K. Miyake, A. Miyake, H. Harima, K. Ishida, Y. Tokunaga, S. Fujimori, F. Hardy, C. Meingast, W. Knafo and A. Zheludev for fruitful discussion.
This work was supported by ERC starting grant (NewHeavyFermion), ANR (FRESCO) and KAKENHI (JP19H00646, JP20K20889, JP20H00130, JP20KK0061, JP18H05227, JP18H01178, JP20H05159, JP19K21840), GIMRT (20H0406), and ICC-IMR.

%\bibliographystyle{jpsj}
%\bibliography{bibbase}
%%%%%%%%%%%%%%%%%%%%%%%%%%%%%%%%%%%%%%%

%============================================
\section*{Supplement}

\subsection*{Experimental}
For the magnetic susceptibility and the magnetization measurements under pressure, a single crystal with the approximate dimension, $1.2\times 1.5\times 1.0\,{\rm mm}$ for $a$, $b$, $c$-axes was selected, 
and the same crystal has been used for the different measurements under pressure and magnetic field.
The mass of the sample is $8.42\,{\rm mg}$, which is large enough compared to the background signal from the pressure cell.
The magnetization and magnetic susceptibility measurements under pressure were performed in a commercial SQUID magnetometer at low temperatures down to $2\,{\rm K}$ and at high fields up to $5\,{\rm T}$, using a CuBe piston cylinder cell with an inner diameter of $2.5\,{\rm mm}$ and the outer diameter of $8.5\,{\rm mm}$. 
Daphne 7373 was used as a pressure transmitting medium.
The pressure was first calibrated in order to know the relation between pressure and force load by detecting the superconducting transition of Pb. 
After checking the reproducibility of pressure as a function of force load, the sample was inserted in a pressure cell without Pb.
The background signal from the pressure cell was subtracted using the data at ambient pressure with/without the pressure cell.

\subsection*{Anisotropy of magnetic susceptibility and inverse magnetic susceptibility under pressure}
\begin{figure}[tbh]
\begin{center}
\includegraphics[width= 0.9\hsize,clip]{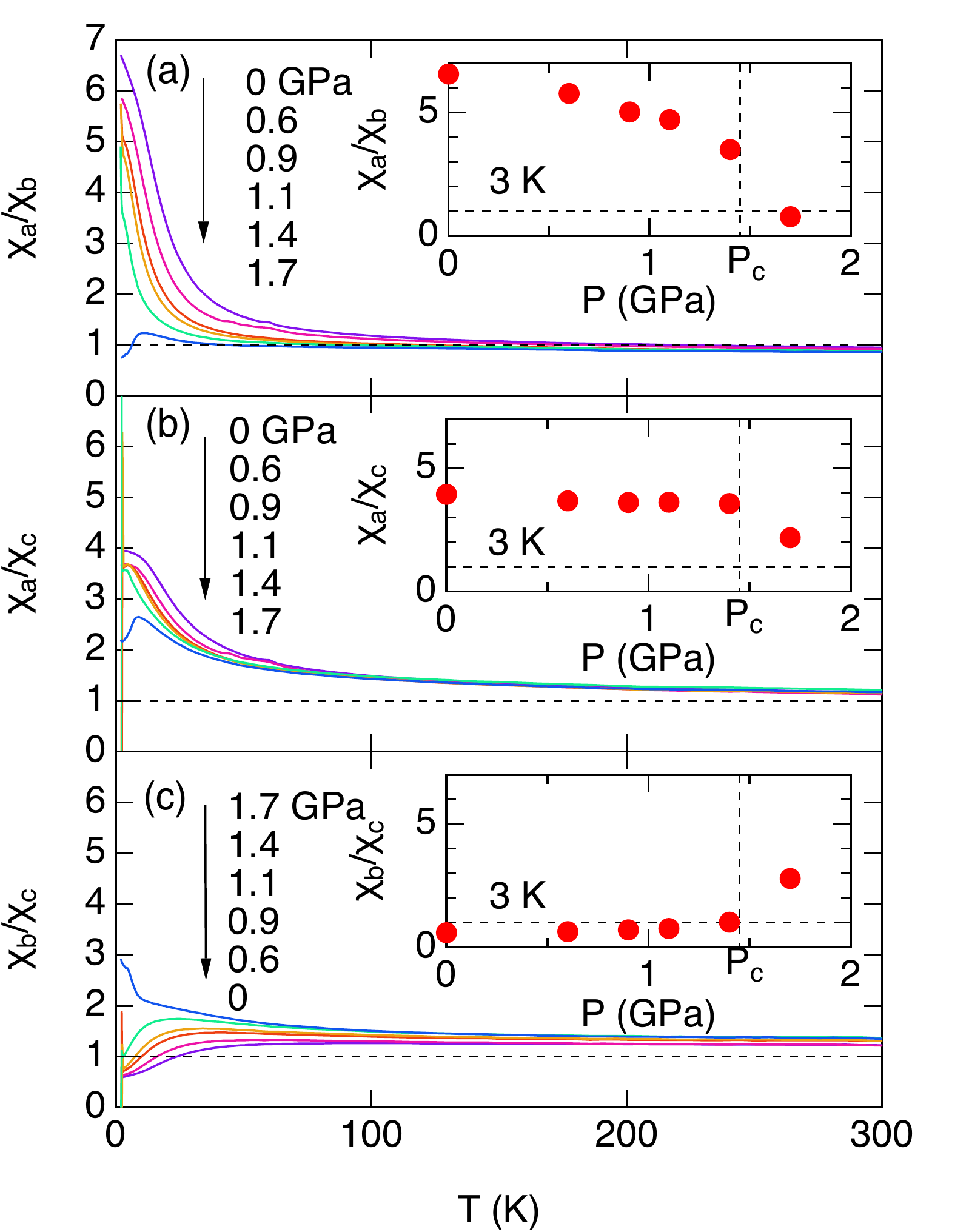}
\end{center}
\caption{Temperature dependence magnetic anisotropy at different pressures. Panels (a), (b), (c) show the ratio of magnetic susceptibilities, $\chi_a/\chi_b$, $\chi_a/\chi_c$ and $\chi_b/\chi_c$, respectively.
The insets in panel (a), (b) and (c) show the pressure dependence of $\chi_a/\chi_b$, $\chi_a/\chi_c$ and $\chi_b/\chi_c$ at $3\,{\rm K}$, respectively.}
\label{fig:sus_anisotropy_suppl}
\end{figure}

Figure~\ref{fig:sus_anisotropy_suppl} shows the temperature dependence of $\chi_a/\chi_b$, $\chi_a/\chi_c$ and $\chi_b/\chi_c$ at different pressures.
At high temperatures, the anisotropies $\chi_a/\chi_b$, $\chi_a/\chi_c$ are close to 1 for all the pressures, indicating rather isotropic magnetic properties.
The magnetic anisotropy is, however, remarkable at low temperatures, and reveals a different pressure response between $\chi_a/\chi_b$, $\chi_a/\chi_c$.
$\chi_a/\chi_b$ at low temperatures decreases rapidly with pressure, getting close to 1.
$\chi_a/\chi_c$ at low temperatures also decreases with pressure, but retaining the value higher than 2.
$\chi_b/\chi_c$ at low temperatures is close to 1 at ambient pressure, indicating both are hard axes compared to the $a$-axis.
Applying the pressure, $\chi_b/\chi_c$ increases slightly at low temperatures.
These results indicate that the magnetic anisotropy becomes small with pressure, retaining the hard magnetization $c$-axis.
%========================================================================================
%========================================================================================

Figure~\ref{fig:inv_sus_suppl} show the temperature dependence of the inverse magnetic susceptibilities for $H\parallel a$, $b$ and $c$-axes at different pressures. 
All the inverse magnetic susceptibilities show linear temperature dependence,
revealing the Curie-Weiss behavior. 
The effective moments and the Weiss constants are extracted from the linear fitting between $150$ and $300\,{\rm K}$, as described in the main text.
%========================================================================================
\begin{fullfigure}[tbh]
\begin{center}
\includegraphics[width=0.8\hsize,clip]{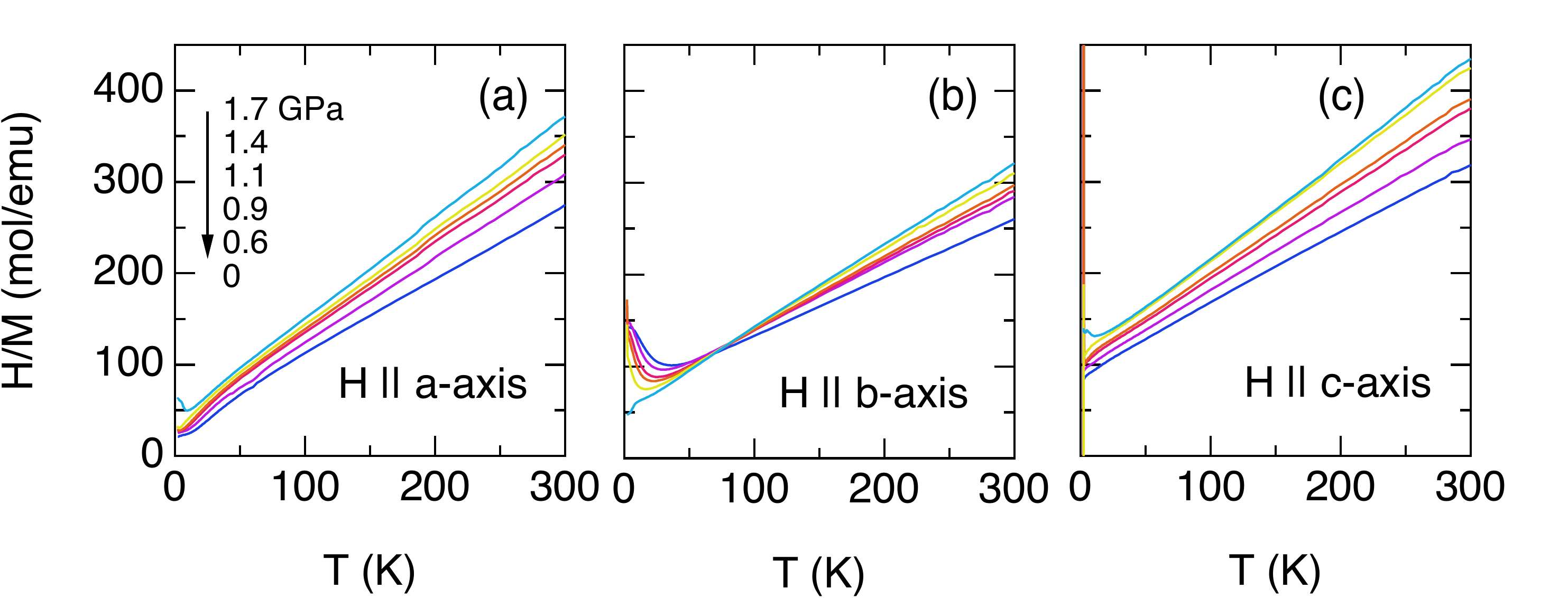}
\end{center}
\caption{Temperature dependence of the inverse susceptibilities for $H\parallel a$, $b$ and $c$-axes at $1\,{\rm T}$ under pressures at 0, 0.6, 0.9, 1.1, 1.4 and $1.7\,{\rm GPa}$.}
\label{fig:inv_sus_suppl}
\end{fullfigure}
%========================================================================================

%---------------------------------------------------------------------------------
\subsection*{First-principles calculation under pressure and discussion on electronic state and magnetism}

\begin{figure}[tbh]
\begin{center}
\includegraphics[width= 0.9\hsize,clip]{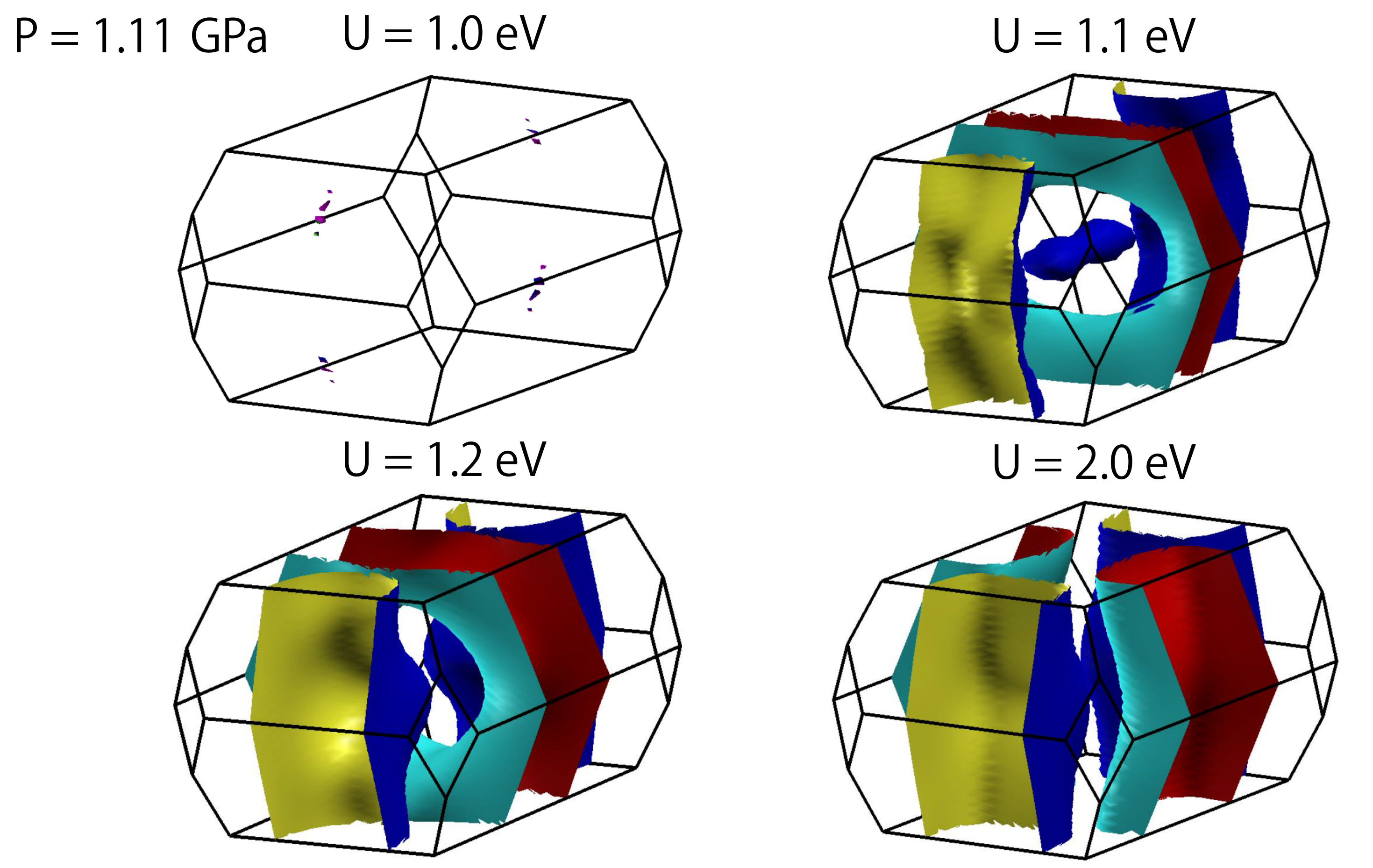}
\end{center}
\caption{Fermi surfaces of UTe$_2$ at $P=1.11$ GPa calculated by the GGA+$U$ moethd for $U=1.0$, $1.1$, $1.2$, and $2.0$ eV.}
\label{fig:FS_p_1.11GPa}
\end{figure}

We show the results of density functional theory plus Hubbard $U$ calculations with the experimentally determined lattice constants $a$, $b$, and $c$ at $P = 1.11$ GPa. The calculations are carried out by using the full-potential linearized augmented plane wave+local orbitals method within the generalized gradient approximation in the \textsc{wien}2k package \cite{Blaha}.
Figure \ref{fig:FS_p_1.11GPa} shows the obtained Fermi surfaces.
An insulator-metal transition occurs at $U=1.0$ eV, and the Fermi surfaces drastically change with increasing $U$ similarly to the case at ambient pressure \cite{Ishizuka2019}. 
%The characteristics of the Fermi surfaces are also similar to those at ambient pressure; 
The electron and hole sheets for $U = 2.0$ eV show a two-dimensional structure, while an electron Fermi surface shows a three-dimensional structure for $U = 1.2$ eV. These Fermi surfaces resemble those obtained by previous calculations at ambient pressure.
%The same feature can be seen at $U=2.0$ eV.
On the other hand, the difference can be found at $U=1.1$ eV. A new hole pocket appears at the $\Gamma$-point. It implies a Lifshitz transition %close to the insulator-metal transition 
under pressure.

The Lifshitz transition may play a role in the pressure dependence of magnetic properties. 
We need further calculations to discuss the magnetic fluctuations and superconductivity based on the obtained band structure. We left them for future study.
Our model calculation \cite{Ish21} found that the magnetic susceptibility with strong Ising-type anisotropy along the $a$-axis is suppressed under pressure and changes to an isotropic one in good agreement with experimental data in the main text. The change in magnetic fluctuations should affect the superconducting symmetry under pressure.

%========================================================================================
%========================================================================================

%\begin{thebibliography}{9}
%\end{thebibliography}
%

\end{document}